\documentclass[preprint]{jpsj3}
\usepackage{color}

\title{Infinitely Multiple Steps in Magnetization of Ferro- and Antiferromagnetic Ising Models with Frustration on a Diamond Hierarchical Lattice}
\author{Yuhei {\sc Hirose}, Akihide  {\sc Oguchi}, and Yoshiyuki {\sc Fukumoto}}
\inst{Department of Physics, Faculty of Science and Technology,\\ Tokyo University of Science, Noda, Chiba 278-8510}
\recdate{\today}
\abst{
Magnetizations of ferro- and antiferromagnetic Ising models with frustration on diamond hierarchical lattices are exactly obtained at zero temperature.
For the zero-field classical spin-liquid phase found in [Kobayashi {\it et al}, J. Phys. Soc. Jpn. 78, 074004 (2009) ], for which frustrating interactions play an important role,
an infinitely small applied magnetic field can induce an infinitely small magnetization, despite classical Ising models that have discrete energy levels.
In antiferromagnetic systems, the magnetization cannot saturate under finite magnetic fields 
owing to the competition between the unfrustrating antiferromagnetic interaction and the Zeeman interaction and an intrinsic long-range nature  of hierarchical lattices.
}

\begin{document}
\maketitle
\section{Introduction}\label{sec:1}
Multistep magnetization phenomena have been experimentally observed in a few frustrated spin systems. 
In particular, $\rm{SrCu}_{2} (\rm{BO}_{3})_{2}$ on the Shastry-Sutherland lattice has attracted  wide attentions experimentally and theoretically.\cite{Kageyama1997,Maignan2000,Kageyama1999,Miyahara1999,Miyahara2003,Fukumoto2001,Liu2007}
Thus far, it is considered that  magnetization plateaus in this system are caused by the orthogonal dimer structure, which tends to suppress the propagation of triplet dimers.
From the viewpoint of the perturbation theory, interdimer couplings lead to effective long-range interactions between triplet dimers. 
Recently multistep magnetizations have  also been observed in TmB$_{4}$ which  seems to be  the classical Ising system rather than the Heisenberg system.\cite{Siemensmeyer2008}
Huang $\textit{et al.}$ have shown the multistep magnetization of the Ising model on Shastry-Sutherland lattices with a long-range interaction, namely, dipole-dipole interaction.\cite{Huang2012} 
Bak and Bruinsma showed that the one-dimensional Ising model with long-range antiferromagnetic interactions exhibits a complete devil's  staircase, i.e., infinite multiple steps between any two steps.\cite{Bak1982}
In this work, we present another exactly solvable model that shows interesting multistep magnetization phenomena.

Hierarchical lattice models can be exactly solved by the renormalization group method.\cite{rf:1}
Since Berker and Ostlund proposed a hierarchical model related to the renormalization group method,\cite{rf:1,rf:2}
many different models on hierarchical lattices have been proposed and studied.\cite{rf:3,rf:4,rf:5,rf:6,rf:7,rf:8}
Our diamond hierarchical lattice has vertices whose coordination numbers increase whenever the stage goes up.  
Therefore we can regard this lattice behavior as long-range interactions.
In this study, we exactly calculate the zero-temperature magnetization of the ferro- and antiferromagnetic Ising models on the diamond hierarchical lattice, and show that there are several types of infinitely multiple-step (IMS) structure.

The thermodynamic behavior of the frustrated Ising model on the diamond hierarchical lattice has been exactly studied without external fields.\cite{Kobayashi2009} 
The building block of our hierarchical lattice is a diamond unit with four nearest-neighbor  ferromagnetic or antiferromagnetic (AF) bonds and one AF diagonal bond.
It was shown that there exist three types of ground state as a function of strength of the frustrating AF diagonal bond:
ferromagnetic or antiferromagnetic long-range ordered phase, classical spin-liquid phase with highly developed short-range order, and paramagnetic phase.\cite{Kobayashi2009} 
In the paramagnetic phase, which has a vanishing ferro- or antiferromagnetic correlation function, we have almost independent $(\uparrow,\downarrow)$ pairs formed by AF diagonal bonds.
This phase has residual entropy, but the Zeeman interaction is just constant in this degenerate manifold.
It is the classical spin liquid phase that has residual entropy, which can be resolved by the Zeeman interaction.
Thus, it is very interesting to study the effect of the magnetic field on this spin liquid phase.
As a result, we have the IMS structure around $(h,m)=(0,0)$ in the magnetization curve, 
where $m\in [0,1]$ represents the magnetization per site and $h\in [0,\infty)$ the magnetic field.
In other words, if a small magnetic field is applied to the spin liquid phase, then we have an induced magnetization despite classical Ising models.
As for the paramagnetic phase, small magnetic fields cannot induce the magnetization, $m(h<h_{\rm c})=0$.
Unexpectedly, our calculation indicates that there exists an IMS structure around $(h,m)=(h_{\rm c},\frac{3}{16})$.
Also, in the antiferromagnetic case, even if the AF diagonal coupling vanishes, we find that the magnetization cannot be saturated under finite magnetic fields, i.e., with the IMS structure around $(h,m)=(\infty,1)$.

This paper is organized as follows.
The diamond hierarchical lattice and frustrated Ising model are described in sect.2.
We study both the antiferromagnetic and ferromagnetic cases.
The magnetization curve in the antiferromagnetic case is studied in sect.3, together with the description of our formulation.
In sect.4, we present the magnetization curve in the ferromagnetic case.
In sect.5, we summarize the results obtained in this study.

\section{Lattice and Antiferromagnetic Hamiltonian}\label{sec:2}

\begin{figure}[b]
\begin{center}
\includegraphics[width=.70\linewidth]{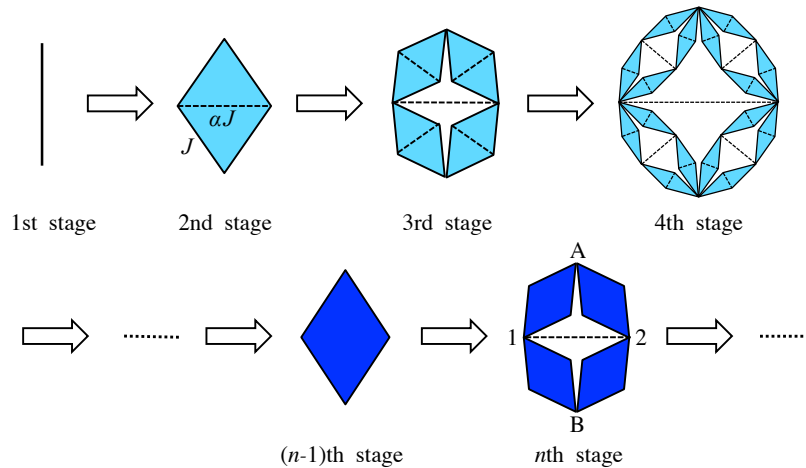}
\end{center}
\caption{
(Color online) Construction of diamond hierarchical lattice.
The indices A and B are assigned to two $(2^{n-1}s)$-type vertices and 1 and 2 are assigned to two $(2^{n-1}s+d)$-type vertices, which are used in Eqs. (3) and (5). }
\label{fig:1}
\end{figure}

We consider the frustrated Ising model on diamond hierarchical lattices.
The diamond hierarchical lattice is constructed by infinite iteration procedures, which are shown in Fig.~{\ref{fig:1}}.

Starting with a single bond expressed by the solid line, which we call the first stage, at the second stage, 
this single bond is replaced by the diamond lattice with four solid lines and a dotted line.
To obtain the third stage, each solid line in the second stage is replaced by the diamond lattice, and the dotted line is left untouched. 
The $n$-stage lattice is constructed by replacing each solid line in the $(n-1)$-stage lattice with the diamond lattice or each solid line in the second-stage lattice with the $(n-1)$-stage lattice. The dotted line is always left untouched.
Thus the $n$-stage lattice has the site number $N_n=\frac{2}{3}(4^{n-1}+2)$, the solid line number $4^{n-1}$, and the dotted line number $\frac{1}{3}(4^{n-1}-1)$. 
There are $2 \times 4^{n-2} $ vertices composed of two solid lines and one dotted line, which we call the $(2s+d)$-type vertex below.
Generally, the number of $(2^{i}s+d)$-type vertices ($i=1, 2, \cdots n-1$) is $2 \times 4^{n-1-i}$.
In addition, there are two $(2^{n-1}s)$-type vertices, which are denoted by A and B in Fig.~{\ref{fig:1}}.

To define the Ising model on each diamond-hierarchical lattice, we place Ising spins on each vertex. 
Spins on both ends of the solid line and of the dotted line couple with each other antiferromagnetically, whose coupling parameters are respectively given by $J$ and  $\alpha J$. 
The Hamiltonian can be written as
\begin{equation} 
   {\cal H}=J\sum_{\langle i,j\rangle} \sigma_i \sigma_j +\alpha J \sum_{\langle\!\langle i,j\rangle\!\rangle}\sigma_i \sigma_j -H\sum_{i}\sigma_{i},
\end{equation}
where the first sum runs over all pairs of nearest neighbors on solid bonds and the second sum runs over all pairs on dotted bonds, and $\sigma_i=\pm 1$.

The diamond unit, which is the same as the second-stage lattice, for example, has the most frustrated ground state at $\alpha=2$. 
The entropy per site of the ground state is $s=\frac{1}{4}\ln 2$ for $\alpha < 2$, $s=\frac{1}{4}\ln 10$ at $\alpha=2$, and $s=\frac{3}{4}\ln 2$ for $\alpha > 2 $. 
Thus, we expect that frustration "$\alpha$" leads to interesting magnetic behavior.

\section{Magnetization of Antiferromagnet with Frustration}\label{sec:3}

The partition function of the second-stage lattice is given by
\begin{equation}
   Z_2=\sum_{\sigma_A,\sigma_B}e^{L(\sigma_A+\sigma_B)}\langle\sigma_{A}|P_{2}|\sigma_{B}\rangle, 
\end{equation}
where
\begin{equation}
   \langle\sigma_A|P_{2}|\sigma_B\rangle=\sum_{\sigma_1,\sigma_2}\exp[-K(\sigma_1+\sigma_2)(\sigma_A+\sigma_B)-B\sigma_1\sigma_2+L(\sigma_{1}+\sigma_{2})]
\label{eq:3}
\end{equation} 
with $K=J/k_{\rm{B}}T$, $B=\alpha J/k_{\rm{B}}T$, and $L=H/k_{\rm{B}}T$.
The partition function of the third-stage lattice is given by
\begin{equation}
    Z_3=\sum_{\sigma_A,\sigma_B}e^{L(\sigma_A+\sigma_B)}\langle\sigma_{A}|P_{3}|\sigma_{B}\rangle,
\end{equation}
where
\begin{equation}
   \langle\sigma_{A}|P_{3}|\sigma_{B}\rangle=\sum_{\sigma_1,\sigma_2}e^{L(\sigma_1+\sigma_2)-B\sigma_1\sigma_2}
   \langle\sigma_A|P_{2}|\sigma_1\rangle\langle\sigma_1|P_{2}|\sigma_B\rangle\langle\sigma_B|P_2|\sigma_2\rangle\langle\sigma_2|P_2|\sigma_A\rangle.
\end{equation}
Adopting similar procedures repeatedly, we have the partition function at the $n$-stage lattice and recurrence formulas:
\begin{equation} 
Z_n=e^{2L}a_{n}+2b_{n}+e^{-2L}c_{n},
\label{eq:5}
\end{equation}
where
\begin{eqnarray} 
   a_{n}&&\hspace{-8mm}\equiv \langle 1|P_{n}|1\rangle=e^{2L-B}a_{n-1}^{4}+2e^{B}a_{n-1}^{2}b_{n-1}^{2}+e^{-2L-B}b_{n-1}^{4},\nonumber\\
   b_{n}&&\hspace{-8mm}\equiv \langle 1|P_{n}|-1\rangle=\langle -1|P_{n}|1\rangle\nonumber\\
           &&\hspace{-8mm}=e^{2L-B}a_{n-1}^{2}b_{n-1}^{2}+2e^{B}a_{n-1}c_{n-1}b_{n-1}^{2}+e^{-2L-B}b_{n-1}^{2}c_{n-1}^{2},
\label{eq:6}
                 \\
   c_{n}&&\hspace{-8mm}\equiv \langle -1|P_{n}|-1\rangle=e^{2L-B}b_{n-1}^{4}+2e^{B}b_{n-1}^{2}c_{n-1}^{2}+e^{-2L-B}c_{n-1}^{4}.\nonumber
\end{eqnarray}
Thus, obtaining the $n$-stage magnetization is equivalent to obtaining $a_{n}$, $b_{n}$, and $c_{n}$ by using repeatedly the recursion formulas with initial values of $a_{2}$, $b_{2}$, and $c_{2}$. 
We, however, face difficulties; as $n$ is increased, the number of terms appearing in the partition function increases like $3^{4^{n-2}}$. 

\subsection{Case of $2K < B \le 3K$}

We begin with the case of a strong AF diagonal coupling, $2K<B$,
for which the paramagnetic phase is stabilized at zero magnetic field.\cite{Kobayashi2009} 
In the course of calculations, we encounter an additional condition, $B \le 3K$, which is used in Eq.~({\ref{eq:13}}). 
We restrict ourselves to the absolute zero temperature.
In the limit $T\rightarrow 0$, we only consider the largest term in $a_{n}$, $b_{n}$, and $c_{n}$.
Fortunately, after repeating  the recursion formula several times, and choosing the largest term, 
we can find simple functional relations between $a_{n}$, $b_{n}$, and $c_{n}$ depending on the intensity of the magnetic field as shown in Table~{\ref{table:1}}. 
Henceforth for brevity, we denote $K$, $B$, and $L$ for $T\rightarrow 0$ instead of $J$, $\alpha J$, and $H$.

\begin{table}[b]
\centering
\caption{Most dominant terms in $a_2$, $b_2$, $c_2$, $a_3$, $b_3$, and $c_3$
as a function of magnetic field $L$ for $B\in(2K,3K]$.}
\label{table:1}
\begin{tabular}{@{\hspace{\tabcolsep}\extracolsep{\fill}}c|ccccccc}
\hline
$L$ &  $[0,B-2K)$ & \vline & $[B-2K,B)$ & \vline & $[B,2K+B)$ & \vline & $[2K+B,\infty)$ \\ \hline
$a_{2}$ & $2e^{B}$ & & $2e^{B}$ & & $2e^{B}$ & \vline  & $e^{2L-B-4K}$ \\ \hline
$b_{2}$ & $2e^{B}$ & & $2e^{B}$ & \vline & $e^{2L-B}$ & & $e^{2L-B}$  \\  \hline
$c_{2}$ & $2e^{B}$ &   \vline   & $e^{2L-B+4K}$&& $e^{2L-B+4K}$&&$e^{2L-B+4K}$ \\ \hline
$a_{3}$ & $2^{5}e^{5B}$&& $2^{5}e^{5B}$ & \vline & $2^{3}e^{4L+B}$ & \vline&\\ \hline       
$b_{3}$ & $2^{5}e^{5B}$ & \vline & $2^{4}e^{2L+3B+4K}$ &\vline& $2^{2}e^{6L-B+4K}$ &\vline   \\ \hline
$c_{3}$ & $2^{5}e^{5B}$ & \vline & $2^{3}e^{4L+B+8K}$ &\vline& $2e^{8L-3B+8K}$ &\vline      \\ \hline        
\end{tabular}
\end{table}

From Table~{\ref{table:1}}, we obtain 
$a_{2}=e^{2L-B-4K}$, $b_{2}=e^{\lambda}a_{2}$, and $c_{2}=e^{2\lambda}a_{2}$ for $L \ge 2K+B$, where $\lambda=4K$;
for $B \le L <2K+B$, $a_{3}=2^{3}e^{4L+B}$, $b_{3}=e^{\lambda}a_{3}$, and $c_{3}=e^{2\lambda}a_{3}$, where $\lambda=2L-2B+4K-\log 2$;
for $B-2K \le L<B$, $a_{3}=2^{5}e^{5B}$, $b_{3}=e^{\lambda}a_{3}$, and $c_{3}=e^{2\lambda}a_{3}$, where $\lambda=2L-2B+4K-\log 2$, 
and for $L<B-2K$, $a_{3}=b_{3}=c_{3}$.

If $a_{i}$, $b_{i}=e^{\lambda}a_{i}$, and $c_{i}=e^{2\lambda}a_{i}$ are given, then by using  Eqs.~(\ref{eq:5}) and (\ref{eq:6}) we obtain the partition function of the $n$-stage:
\begin{equation}
   \log Z_n=\log(e^{2L}+2e^{2^{n-i}\lambda}+e^{-2L+2^{n-i+1}\lambda})+4^{n-i}\log a_{i}+\sum_{r=1}^{n-i}4^{n-i-r}\log F(2^{r}\lambda),
\end{equation}
where
\begin{equation}
   F(2^{r}\lambda )=e^{2L-B}+2e^{B+2^{r}\lambda}+e^{-2L-B+2^{r+1}\lambda}.
\label{eq:8}
\end{equation}
The magnetization per site for $n=\infty$ is obtained as 
\begin{equation}
m=\frac{3}{2}\left\{ \frac{1}{4^{i-1}} \frac{\partial}{\partial L}\log a_{i}+\sum_{r=1}^{\infty}\frac{1}{4^{i-1+r}}G( r)\right\},
\label{eq:9}
\end{equation}
where 
\begin{equation}
 G(r)=\frac{\partial}{\partial L}\log F(2^{r}\lambda) .
\label{eq:10}
\end{equation}
This equation is the central equation in this paper.

\begin{figure}[b]
\begin{center}
\includegraphics[width=.75\linewidth]{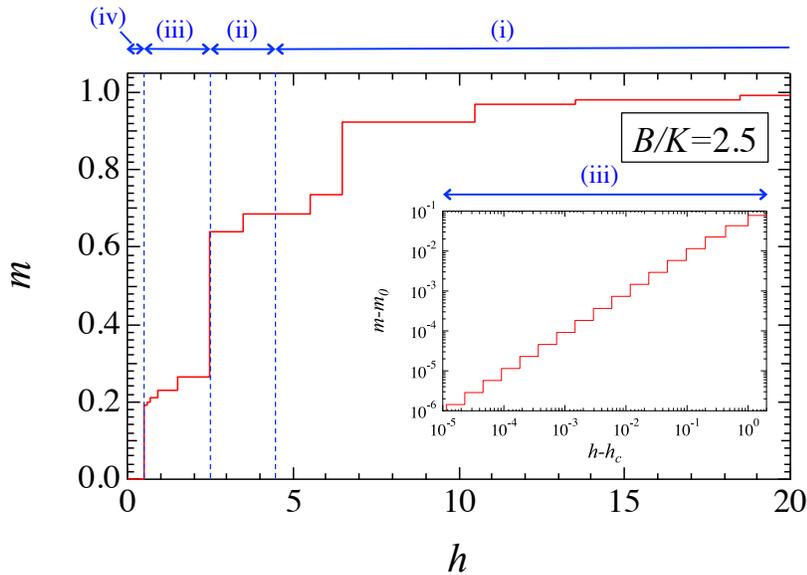}
\caption{(Color online)  Dependence of magnetization per site $m$ on magnetic field 
$h=H/J=L/K$ for $\alpha=B/K=2.5$.
The inset shows an enlarged plot of region (iii), where 
$h_{\rm c}=B/K-2=0.5$ is the extreme left of region (iii) and $m_0=3/16$
is the magnetization at $h=h_{\rm c}+0$. 
 }
\label{fig:2}
\end{center}
\end{figure}

Before proceeding to the calculation process, we summarize the typical features of the resultant magnetization curve for $2K<B\le 3K$.
We show a calculated result in Fig.~\ref{fig:2}, where $B/K=2.5$ is chosen.
The regions indicated in Fig.~\ref{fig:2}, (i) $2K+B\le L<\infty$, (ii) $B\le L<2K+B$, (iii) $B-2K\le L<B$, and (iv) $0\le L<B-2K$, correspond to those in Table \ref{table:1}.
In region (i), we have an infinitely large saturation field. 
Spins on $(2^is+d)$-type vertices flip upwards, whenever a magnetic field is added to $L=2^iK+B$.
In other words, we have the IMS structure around $(m,L)=(1,\infty)$.
In region (iii), there appears another IMS structure around $(m,L)=(\frac{3}{16},L_{\rm c})$,
where $L_{\rm c}=B-2K$ is a critical field, $m(L<L_{\rm c})=0$.
In region (iv), the magnetization vanishes owing to strong AF diagonal bonds.
Also, it can be proved that the height of each magnetization plateau is independent of $B/K\in(2,3]$.

\nobreak

\clearpage

We present the calculation process for regions (i)-(iv) in the following subsections, {\it\ref{sec:3-1-1}}-{\it\ref{sec:3-1-4}}, respectively.

\subsubsection{$2K+B\le L$}\label{sec:3-1-1}

In this magnetic field region, we have $a_{2}=e^{2L-B-4K}$, $b_{2}=e^{\lambda}a_{2}$, and $c_{2}=e^{2\lambda}a_{2}$ with $\lambda=4K$. 
Substituting these into Eq. (\ref{eq:9}), we obtain
\begin{equation}
m=\frac{3}{2}\left\{\frac{2}{4}+\sum_{r=1}^{\infty}\frac{1}{4^{r+1}}G( r)\right\}.
\label{eq:11new}
\end{equation}
Using Eqs. (\ref{eq:8}) and (\ref{eq:10}), and choosing the largest term in each $G(r)$, we obtain
\begin{equation}
G(1)=\begin{cases}
0 & \mbox{for $2K+B \le L < 4K+B$}\\
2 & \mbox{for $4K+B \le L$}
\end{cases},
\end{equation}   
and for $r \ge 2 $
\begin{equation}
G(r)=\begin{cases}
-2 & \mbox{for $L < 2^{r+1}K-B$} \\
0 & \mbox{for $2^{r+1}K-B \le  L < 2^{r+1}K+B$} \\
2 &  \mbox{for $2^{r+1}K+B \le L$}
\end{cases}.
\end{equation}   
The magnetization is obtained as
\begin{equation}
   m(2K+B \le L \le 8K-B)=\frac{3}{2}\left(\frac{2}{4}+\frac{0}{4^{2}}-\frac{2}{4^{3}}-\cdots \right) =\frac{11}{16}=0.6875,
\label{eq:13}
\end{equation}
where we have used the fact that the condition $B\leq 3K$ ensures $2K+B\leq 8K-B$.
Similarly, we have
\begin{equation}
   m(8K-B \le L < 4K+B)=\frac{3}{2}\left(\frac{2}{4}+\frac{0}{4^{2}}+\frac{0}{4^{3}}-\frac{2}{4^{4}}-\cdots\right)=\frac{47}{64}=0.734375,
\label{eq:14}
\end {equation}
\begin{equation}
   m(4K+B \le L < 8K+B)=\frac{3}{2}\left(\frac{2}{4}+\frac{2}{4^{2}}-\frac{0}{4^{3}}-\frac{2}{4^4}-\cdots\right) =\frac{59}{64}=0.921875.
\label{eq:15}
\end{equation}
For $r\ge 3$, we have
\begin{eqnarray}
   m(2^{r}K+B \le L  < 2^{r+1}K-B)&&\hspace{-6mm}=\frac{3}{2}\left(\frac{2}{4}+\frac{2}{4^{2}}+\cdots +\frac{2}{4^{r}}-\frac{2}{4^{r+1}}-\frac{2}{4^{r+2}}-\cdots\right)
   \nonumber\\ 
   &&\hspace{-6mm}=1-\frac{2}{4^{r}},
\label{eq:16}
\end{eqnarray}
and 
\begin{eqnarray}
   m(2^{r+1}K-B \le L < 2^{r+1}K+B)&&\hspace{-6mm}=\frac{3}{2}\left(\frac{2}{4}+\frac{2}{4^{2}}+\cdots +\frac{2}{4^{r}}-\frac{0}{4^{r+1}}-\frac{2}{4^{r+2}}-\cdots \right)
   \nonumber\\ 
   &&\hspace{-6mm}=1-\frac{1}{4^{r}}-\frac{1}{4^{r+1}}.
\label{eq:17}
\end{eqnarray}
Some values of the magnetization are given below: 
$m(2^{2}K+B \le L < 2^{3}K+B)=\frac{59}{64}=0.921875$, $m(2^{3}K+B \le L < 2^{4}K-B)=\frac{31}{32}=0.96875$, and $m(2^{4}K-B \le L < 2^{4}K+B)=\frac{251}{256}\simeq0.980469$.

These results show the position of the magnetic field where the magnetization jumps up depends on the magnitudes of $K$ and $B$, 
but its jumping width has universality, which does not depend on these parameters. (This universality holds for the other regions of the magnetic field.)
Also, the magnetization of the frustrated antiferromagnetic diamond hierarchical lattice is not saturated.
For the present high-magnetic-field region, as discussed below, spins on $(2^{r}s+d)$-type vertices flip upwards, whenever magnetic fields are added to $2^{r}K+B$.

\begin{figure}[h]
\begin{center}
\includegraphics[width=.65\linewidth]{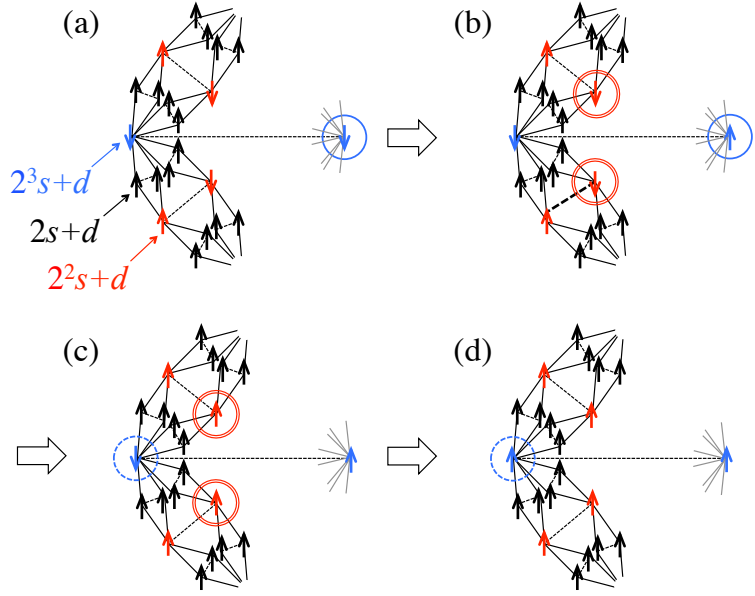}
\end{center}
\caption{(Color)  Spin configurations (a) $\phi_{\rm a}$, (b) $\phi_{\rm b}$, (c) $\phi_{\rm c}$, and (d) $\phi_{\rm d}$.
Flipped spins are marked by circles. }
\label{fig:2p}
\end{figure}

We consider the $n$-stage lattice, for which the total number of vertices is $N_n=\frac{2}{3}(4^{n-1}+2)$.
We denote the magnetization per site with respect to $(2^is+d)$-type vertices as $m_i$.
Noting that the number of $(2^is+d)$-type vertices ($i=1,2,\cdots,n-1$) is $N_n^{(i)}=2\times 4^{n-1-i}$ and that of $(2^{n-1}s)$-type vertices is two,
we can write the total magnetization per site, $m$, in terms of that on each type of vertex as 
\begin{equation}
   m=\frac{1}{N_n}\left[\sum_{i=1}^{n-1}N_n^{(i)}m_i+2m_{\rm s}\right],
\end{equation}
where $m_{\rm s}$ is the magnetization per site for $(2^{n-1}s)$-type vertices.
In the limit $n\rightarrow\infty$, we have
\begin{equation}
   m=\frac{3}{2}\left[\frac{2}{4}m_1+\frac{2}{4^2}m_2+\frac{2}{4^3}m_3+\cdots\right].
\label{eq:m_1}
\end{equation}
We compare Eq.~(\ref{eq:13}) with Eq.~(\ref{eq:m_1}), and assume that the lowest-energy configurations for $L\in(2K+B,8K-B)$ have
\begin{equation}
   m_1=1,\;\;\;m_2=0,\;\;\;m_3=-1,\;\;\;m_4=-1,\;\cdots,
\label{eq:20}
\end{equation}
i.e., all the spins on $(2^1s+d)$-type vertices are up, half of the spins on $(2^2s+d)$-type vertices are up and the other half are down, and all the spins on the other vertices are down.
In the same way, comparing Eqs.~(\ref{eq:14})-(\ref{eq:17}) with Eq. (\ref{eq:m_1}), 
we assume $\{m_i\}$ in the lowest-energy spin configurations for each of the magnetic field regions, as shown in Table \ref{table:1p}.
In Fig. \ref{fig:2p}, we show the lowest-energy spin configurations $\phi_{\rm a}$, $\phi_{\rm b}$, $\phi_{\rm c}$, and $\phi_{\rm d}$ for the first four magnetic field regions.
In Table \ref{table:1pp}, we list the changes in exchange energy, $\Delta E_{\rm{exchange}}$, and Zeeman energy, $\Delta E_{\rm{Zeeman}}$, 
as well as the critical magnetic fields $L_c$, which are determined by $\Delta E_{\rm{exchange}}+\Delta E_{\rm{Zeeman}}=0$, 
with respect to $\phi_{\rm a}\rightarrow\phi_{\rm b}$, $\phi_{\rm b}\rightarrow\phi_{\rm c}$, and $\phi_{\rm c}\rightarrow\phi_{\rm d}$.
The resultant critical fields are in agreement with the boundary values for Eqs.~(\ref{eq:13})-(\ref{eq:15}) and~(\ref{eq:16}) with $r=3$.
Also, it is easy to show that the other critical fields can be reproduced by the same procedure.
This fact proves that our assumptions for $\{m_i\}$ are true.
Thus, we conclude that spins on $(2^{r}s+d)$-type vertices flip upwards, whenever magnetic fields are added to $2^{r}K+B$.

\begin{table}[t]
\centering
\caption{Assumptions for $\{m_i\}$ as a function of $L$.}
\label{table:1p}
\begin{tabular}{c|c|c|c|c|c|c} \hline
$L$                      & $m_1$ & $m_2$ & $m_3$ & $m_4$ & $m_5$ & $m_6$ \\ \hline
$[2K+B,8K-B)$    &    1       &      0     &     -1     &     -1      &     -1     &    -1      \\
$[8K-B,4K+B)$    &    1       &      0     &      0     &      -1     &     -1     &    -1      \\
$[4K+B,8K+B)$   &    1       &      1     &      0     &      -1     &     -1     &    -1      \\ \hline
$[8K+B,16K-B)$  &    1       &      1     &      1     &      -1     &     -1     &    -1      \\
$[16K-B,16K+B)$&    1       &      1     &      1     &      0      &     -1     &    -1       \\ \hline
$[16K+B,32K-B)$&    1       &      1     &      1     &      1      &     -1     &    -1       \\
$[32K-B,32K+B)$&    1       &      1     &      1     &      1      &      0     &    -1       \\ \hline
\end{tabular}
\end{table}

\begin{table}[t]
\centering
\caption{Changes in exchange and Zeeman energies and critical fields for transitions 
$\phi_{\rm a}\rightarrow\phi_{\rm b}$, $\phi_{\rm b}\rightarrow\phi_{\rm c}$, and $\phi_{\rm c}\rightarrow\phi_{\rm d}$.}
\label{table:1pp}
\begin{tabular}{c|c|c|c} \hline
                                             &         $a\to b$                   &  $b\to c$                            & $c\to d$                                \\ \hline
$\Delta E_{\rm{exchange}}$& $(8K-B)N_{\infty}^{(3)}$   &  $(4K+B)N_{\infty}^{(2)}$  & $(8K+B)N_{\infty}^{(3)}$     \\ \hline
$\Delta E_{\rm{Zeeman}}$  &  $-L N_{\infty}^{(3)}$         &  $-L N_{\infty}^{(2)}$         & $-L N_{\infty}^{(3)}$             \\ \hline
$L_{\rm{c}}$                        &  $8K-B$                            &  $4K+B$                            & $8K+B$                              \\ \hline
\end{tabular}
\end{table}

Comparing Eq.~(\ref{eq:m_1}) with Eq.~(\ref{eq:11new}), we note that the above-mentioned discussion suggests that 
\begin{equation}
m_1=\frac{1}{2} \frac{\partial}{\partial L}\log{a_2},\;\;\;m_r=\frac{1}{2} G(r-1)\;\;\;\mbox{for $r \ge 2$}
\label{eq:22new}
\end{equation}
hold in the present high-field region.

\subsubsection{$B \le L<2K+B$}
 
From Table~{\ref{table:1}}, we have $a_{3}=2^{3}e^{4L+B}$, $b_{3}=e^{\lambda}a_{3}$, and $c_{3}=e^{2\lambda}a_{3}$, where $\lambda=2L-2B+4K-\log2$. 
If we define 
\begin{equation}
L_{r}=\frac{(2^{r}+1)B-2^{r+1}K}{2^{r}-1},
\label{eq:19}
\end{equation}
we find $2K+B >L_{1}=3B-4K > B>L_{2} >L_{3} \cdots $. 
Using these relations, we get when $2K+B>L \ge L_{1}=3B-4K$
\begin{equation}
   G(r)=-2+2^{r+2},
\end{equation}
and when $B \le L<3B-4K$,
\begin{equation}
   G(r)=\begin{cases}
   2^{2} & \mbox{for $r=1$}\\
   2^{r+2}-2  &  \mbox{for $r \ge 2$}
   \end{cases}.
\end{equation}
Noting  $i=3$ in Eq.~(\ref{eq:9}) and substituting these values into Eqs.~(\ref{eq:9}) and (\ref{eq:10}), we get 
\begin{eqnarray}
   m(3B-4K \le L< B+2K)&&\hspace{-6mm}=\frac{3}{2}\left(\frac{1}{4}+\frac{2^{3}-2}{4^{3}}+\frac{2^{4}-2}{4^{4}}+\frac{2^{5}-2}{4^{5}}+\cdots\right)
   \nonumber\\
   &&\hspace{-6mm}= \frac{44}{64}=0.6875,
\label{eq:26new}   
\end{eqnarray}
and 
\begin{eqnarray}
   m(B \le L  < 3B-4K)&&\hspace{-6mm}=\frac{3}{2}\left(\frac{1}{4}+\frac{2^{2}}{4^{3}}+\frac{2^{4}-2}{4^{4}}+\frac{2^{5}-2}{4^{5}}+\cdots\right)
   \nonumber\\
   &&\hspace{-6mm}=\frac{41}{64}=0.640625.
\label{eq:27new}   
\end{eqnarray}

Now, we try to identify spin configurations, following the same procedure as in the previous subsection.
We first note that the comparison between Eqs.~({\ref{eq:m_1}}) and ({\ref{eq:26new}}) does not lead to any meaningful assumption for $\{m_i\}$.
Thus, we rewrite Eq.~({\ref{eq:26new}}) as
\begin{eqnarray}
   m(3B-4K \le L< B+2K)&&\hspace{-6mm}=\frac{3}{2}\left\{\left(\frac{1}{4}+\frac{2^{3}}{4^{3}}+\frac{2^{4}}{4^{4}}+\cdots\right)
   -\frac{0}{4^{2}}-\frac{2}{4^{3}}-\frac{2}{4^{4}}-\cdots\right\}
   \nonumber\\
   &&\hspace{-6mm}=\frac{3}{2}\left(\frac{2}{4}-\frac{0}{4^{2}}-\frac{2}{4^{3}}-\frac{2}{4^{4}}-\cdots\right).
\end{eqnarray}
This expression is the same as Eq.~(\ref{eq:13}) and leads to the assumption in Eq.~(\ref{eq:20}), as expected.
In a similar way, Eq.~(\ref{eq:27new}) can be rewritten as 
\begin{eqnarray}
   m(B\le L< 3B-4K)&&\hspace{-6mm}=
   \frac{3}{2}\left\{\left(\frac{1}{4}+\frac{2^{3}}{4^{3}}+\frac{2^{4}}{4^{4}}+\cdots\right)
   -\frac{1}{4^{2}}-\frac{0}{4^{3}}-\frac{2}{4^{4}}-\cdots\right\}
   \nonumber\\
   &&\hspace{-6mm}=\frac{3}{2}\left(\frac{2}{4}-\frac{1}{4^{2}}-\frac{0}{4^{3}}-\frac{2}{4^{4}}-\cdots\right),
\end{eqnarray}
which suggests that
\begin{equation}
   m_1=1,\;\;\;m_2=-\frac{1}{2},\;\;\;m_3=0,\;\;\;m_4=-1,\;\cdots.
\end{equation}
For the transition between the lowest-energy spin configurations for these two assumptions for $\{m_i\}$,
we have $\Delta E_{\rm{exchange}}=(-8K+B)N^{(3)}_{\infty}+(2K-\frac{1}{2}B)N^{(2)}_{\infty}$
and $\Delta E_{\rm{Zeeman}}=-LN^{(3)}_{\infty}+\frac{1}{2}LN^{(2)}_{\infty}$.
However, the condition $\Delta E_{\rm{exchange}}+\Delta E_{\rm{Zeeman}}=0$ gives an unreasonable critical field $L=-B$.
We have not succeeded in identifying spin configurations except for the highest magnetic field region.

\subsubsection{$B-2K \le L < B$}
 
From Table~{\ref{table:1}}, we have $a_{3}=2^{5}e^{5b}$ and $\lambda=2L-2B+4K-\log2$. 
Thus we get
\begin{equation}
G(r)=\begin{cases}
2^{r+1} & L < L_{r} \\
2^{r+2}-2 & L \ge L_{r}  ,
\end{cases}
\end{equation}
where $L_{r}$ is given by Eq.~(\ref{eq:19}). Note that $L_{2} < B$.
Substituting these into Eqs.~(\ref{eq:9}) and (\ref{eq:10}), we obtain
 \begin{equation}
   m(L_{2} \le L  <  B)=\frac{3}{2}\left( \frac{2^{2}}{4^{3}}+\sum_{s=2}^{\infty}\frac{2^{s+2}-2}{4^{s+2}}\right),
\label{eq:25}
\end{equation}
and
\begin{equation}
   m(L_{r+1} \le L  <  L_{r})=\frac{3}{2}\left( \sum_{s=1}^{r}\frac{2^{s+1}}{4^{s+2}}+\sum_{s=r+1}^{\infty}\frac{2^{s+2}-2}{4^{s+2}}\right).
\label{eq:26}
\end{equation}

From Eqs.~(\ref{eq:25}) and (\ref{eq:26}), we get  $m(L=B \le L_{1})=\frac{17}{64}=0.265625$ and $m(L=L_{\infty}=B-2K)=\frac{3}{16}=0.1875$.
We observe that the IMS structure appears and goes down towards $\frac{3}{16}$ from $\frac{17}{64}$. 
Values of the magnetization from above are given by $\frac{17}{64}$, $\frac{59}{256}$, $\frac{215}{1024}$, $\frac{815}{4096}$, $\cdots$, $\frac{3}{16}$. 
The width of a plateau becomes narrower as $L\rightarrow L_{\infty}+0$, or $h\rightarrow h_{c}+0$, which is clealy seen in the inset of Fig.~{\ref{fig:2}}.
In our hierarchical lattices, many solid lines entering into a vertex  play a role as the long-range interactions.

\subsubsection{$0 \le L < B-2K$}\label{sec:3-1-4}

From Table~{\ref{table:1}}, we have $a_{3}=2^{5}e^{5B} $ and $\lambda=0$.
Substituting these into Eqs.~(\ref{eq:9}) and (\ref{eq:10}), we obtain $m=0$.
In the present case, the AF diagonal coupling $B$ is much stronger than $K$,
and thus $(\uparrow,\downarrow)$ pairs on $B$-bonds appear in the ground state when $h=L/K=0$.
When a small magnetic field is applied, the AF diagonal coupling $B$ prevents magnetization to be induced.

\subsection{Case of $\frac{3K}{2} \le B < 2K$}

We turn to the case of weak AF diagonal coupling, $2K>B$,
for which the zero-field phase is the spin liquid.\cite{Kobayashi2009}
An additional condition $\frac{3K}{2} \le B$ appears in the course of calculations.
In the case of $B < 2K$, we also get the highest term in $a_{2}$, $b_{2}$, and $c_{2}$. The results are shown in Table~{\ref{table:2}}.

\begin{table}[h]
\centering
\caption{Most dominant terms in $a_2$, $b_2$, and $c_2$
as a function of magnetic field $L$ for $B\in[\frac{3}{2}K,2K)$.}
\label{table:2}
\begin{tabular}{@{\hspace{\tabcolsep}\extracolsep{\fill}}c|ccccccc}
\hline
$L$  & $[0,2K-B)$ &\vline& $[2K-B,B)$  &\vline& $[B,2K+B)$ &\vline&  $[2K+B,\infty)$    \\ \hline
$a_{2}$ & $e^{-B-2L+4K}$ & \vline& $2e^{B}$ & & $2e^{B}$ &\vline & $e^{-B+2L-4K}$ \\ \hline
$b_{2}$ & $2e^{B}$               &  & $2e^{B}$ & \vline & $e^{-B+2L}$ && $e^{-B+2L}$ \\ \hline
$c_{2}$ & $e^{-B+2L+4K}$ & & $e^{-B+2L+4K}$ & & $e^{-B+2L+4K}$ & & $e^{-B+2L+4K}$ \\ \hline
\end{tabular}
\end{table}

\begin{figure}[t]
\begin{center}
\includegraphics[width=.75\linewidth]{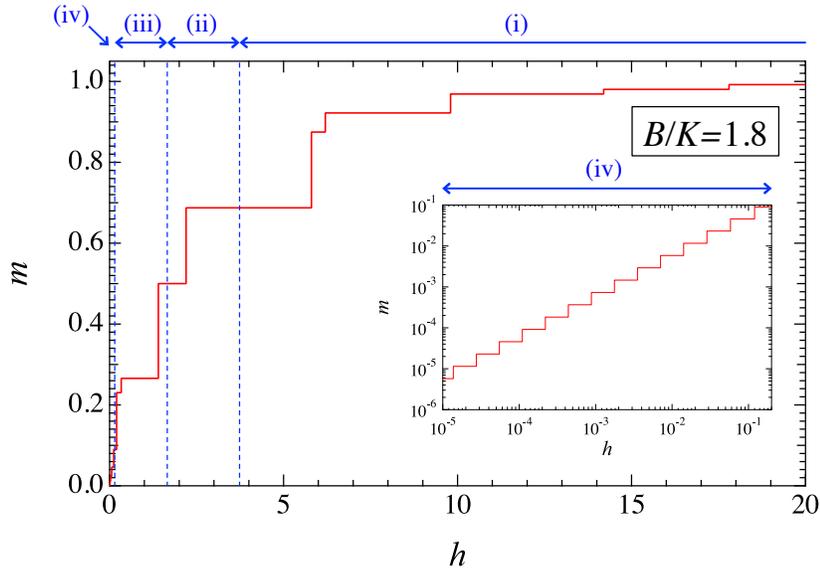}
\end{center}
\caption{(Color online) Dependence of $m$ on $h=L/K$ for $\alpha=B/K=1.8$.
The inset shows an enlarged plot of region (iv), where the
 IMS structure appears.
}
\label{fig:4}
\end{figure}

Here, we describe a typical feature of the magnetization curve for $\frac{3K}{2} \le B < 2K$, before giving the calculation processes.
In Fig. \ref{fig:4}, we show the calculation result for $B/K=1.8$.
The regions indicated in Fig.~\ref{fig:4}, (i) $2K+B\le L<\infty$, (ii) $B\le L<2K+B$, (iii) $2K-B\le L<B$, and (iv) $0\le L<2K-B$, correspond to those in Table \ref{table:2}.
In region (i), we have an infinitely large saturation field or the IMS structure around $(m,L)=(1,\infty)$, which is the same as in the previous case of $2K<B<3K$. 
In region (iii), the total number of magnetization plateaus depends on $B/K$, which differs from the previous case. 
When $B/K=1.8$, we have three steps in region (iii). When $B/K\rightarrow 2$, the total number of steps reaches an infinite value.
In region (iv), the IMS structure around $(m,L)=(0,0)$ appears.
It should be stressed that a small magnetic field can induce a small magnetization, despite a classical Ising system.

We present the calculation process for regions (i)-(iv) in {\it\ref{sec:3-2-1}}-{\it\ref{sec:3-2-4}}, respectively.

\subsubsection{$L \ge 2K+B$}\label{sec:3-2-1}

In this region, we observe that $a_{2}=e^{-B+2L-4K} $ and $b_{2}=e^{\lambda}a_{2}$, and $c_{2}= e^{2\lambda}a_{2}$ with $\lambda=4K$.
Substituting these relations into Eq.~(\ref{eq:9}), we get 
\begin{equation}
   m=\frac{3}{2}\left\{\frac{2}{4}+\sum_{r=1}^{\infty}\frac{1}{4^{r+1}}G( r)\right\}.
\label{eq:27}
\end{equation}
The largest term in each $G(r)$ is given by 
\begin{equation}
G(1)=\begin{cases}
      0 & \mbox{for $2K+B \le L < 4K+B$}\\
      2 & \mbox{for $4K+B\le L$} 
      \end{cases},
\label{eq:28}
\end{equation}
 and  for $r \ge 2$, we get
\begin{equation}
G(r)=\begin{cases}
-2 & \mbox{for $L < 2^{r+1}K-B$} \\
0 & \mbox{for $2^{r+1}K-B \le  L < 2^{r+1}K+B$} \\
2 &  \mbox{for $2^{r+1}K+B \le L$}
\end{cases}.
\label{eq:29}
\end{equation}
The magnetization is obtained from Eqs.~(\ref{eq:27})-(\ref{eq:29}):
\begin{equation}
  m(2K+B \le L <  4K+B)=1-\frac{1}{4}-\frac{1}{4^{2}}=\frac{11}{16}=0.6875,
\end{equation}
and for $r \ge 2$
\begin{equation}
   m(2^{r}K+B \le L <  2^{r+1}K-B) =1-\frac{2}{4^{r}},
\end{equation}
\begin{equation}
   m(2^{r+1}K-B \le L < 2^{r+1}K+B)=1-\frac{1}{4^{r}}-\frac{1}{4^{r+1}}.
\end{equation}

Adopting Eq.~(\ref{eq:22new}) to the present case, we get $\{m_i\}$ shown in Table \ref{table:2p},
which gives correct critical fields.
As the magnetic field increases, half of the spins on $(2^is+d)$-type vertices flip upwards at $L=2^iK-B$ and the other half at $L=2^iK+B$.

\begin{table}[h]
\centering
\caption{Assumptions for $\{m_i\}$ as a function of $L$.}
\label{table:2p}
\begin{tabular}{c|c|c|c|c} \hline
$L$                       & $m_1$ & $m_2$ & $m_3$ & $m_4$   \\ \hline
$[2K+B,4K+B)$    &    1       &      0     &     -1     &     -1       \\
$[4K+B,8K-B)$     &    1       &      1     &     -1     &      -1      \\\hline
$[8K-B,8K+B)$     &    1       &      1     &      0     &      -1      \\
$[8K+B,16K-B)$   &    1       &      1     &      1     &      -1      \\ \hline
\end{tabular}
\end{table}

\subsubsection{$B \le L  <  2K+B$}

In this region, we cannot find the simple relation between $a_2, b_2,$ and $ c_2$ from Table~{\ref{table:2}}. 
We insert $a_{2}$, $b_{2}$, and $c_{2}$ into Eq.~(\ref{eq:6}), and for $4K-B \le L \le 2K+B$ we get the relation between $a_{3}$, $b_{3}$, and $c_{3}$. 
However, for $B \le L \le 4K-B$, we cannot find the relation. 
Therefore, we use Eq.~(\ref{eq:6}), repeat the calculation, and finally obtain Table~{\ref{table:3}}.

\begin{table}[h]
\centering
\caption{
 Most dominant terms in $a_3$, $b_3$, $c_3$, $a_4$, $b_4$, and $c_4$
as a function of magnetic field $L\in[B,2K+B)$ for $B\in[\frac{3}{2}K,2K)$.}
\label{table:3}
\begin{tabular}{c|c|c} \hline
$L$ & $[B,4K-B)$ & $[4K-B,2K+B)$      \\ \hline
$a_{3}$ & \multicolumn{1}{c}{$8e^{B+4L}$} &  $8e^{B+4L}$ \\ \hline
$b_{3}$ & \multicolumn{1}{c}{$4e^{-B+4K+6L}$} & $4e^{-B+4K+6L}$ \\ \hline
$c_{3}$ & $e^{-5B+16K+6L}$ & $2e^{-3B+8K+8L}$ \\ \hline
$a_{4}$ & $256e^{-5B+16K+22L}$ & \\ \hline
$b_{4}$ & $16e^{-13B+40K+22L}$ & \\ \hline
$c_{4}$ & $e^{-21B+64K+22L}$ &   \\ \hline
\end{tabular}
\end{table}

\noindent
(a) $4K-B \le L <  2K+B$

We have $a_{3}=8e^{B+4L}$, $b_{3}=e^{\lambda}a_{3}$, and  $c_{3}=e^{2\lambda}a_{3}$, where $\lambda= -2B+4K+2L-\log 2$. 
Thus, we obtain $G(r) = 2^{r+2}-2$ and the magnetization:
 \begin{equation}
   m(4K-B\le L <  2K+B)= \frac{3}{2}\left(\frac{1}{4}+\sum_{r=1}^{\infty}\frac{2^{r+2}-2}{4^{r+2}}\right)=\frac{11}{16}=0.6875.
\end{equation}

\noindent
(b) $B \le L<4K-B$

In this region, we have $a_{4}=256e^{-5B+16K+22L}$, $b_{4}=e^{\lambda}a_{4}$, and  $c_{4}=e^{2\lambda}a_{4}$, where $\lambda=-8B+24K-4\log 2$. 
Promptly, we have $G(r)=-2$ for $r\ge 1$, 
and the magnetization is 
\begin{equation}
   m(B \le L  <  4K-B)= \frac{3}{2}\left(\frac{22}{4^{3}}-\sum_{r=1}^{\infty}\frac{2}{4^{r+3}}\right)=\frac{1}{2} .
\end{equation}

\subsubsection{$2K-B \le L <  B$}

In this region, Table~{\ref{table:4}} is obtained similarly.

\begin{table}[h]
\centering
\caption{Most dominant terms in $a_3$, $b_3$, $c_3$, $a_4$, $b_4$, and $c_4$
as a function of magnetic field $L\in[2K-B,B)$ for $B\in[\frac{3}{2}K,2K)$.}
\label{table:4}
\begin{tabular}{c|c|c} \hline
$L$ & $[2K-B,3B-4K)$ &$[3B-4K,B)$  \\ \hline
$a_{3}$ & \multicolumn{1}{c}{$32e^{5B}$} & $32e^{5B}$  \\ \hline
$b_{3}$ & \multicolumn{1}{c}{$16e^{3B+4K+2L}$} & $16e^{3B+4K+2L}$ \\ \hline
$c_{3}$ & $8e^{B+8K+4L}$ &  $e^{-5B+16K+6L}$  \\ \hline
$a_{4}$ & &  $16^4 e^{11B+16K+6L}$  \\ \hline
$b_{4}$ & &  $16^2e^{-5B+40K+14L}$ \\ \hline
$c_{4}$ & &  $e^{-21B+64K+22L}$ \\  \hline
\end{tabular}
\end{table}

\noindent
(a) $3B-4K \le L  <  B$

In this region, we have $a_{4}=16^4e^{11B+16K+6L}$, $b_{4}=e^{\lambda}a_{4}$, and $c_{4}=e^{2\lambda}a_{4}$, where $\lambda=-16B+24K+8L-8\log 2$. 
The largest term of $G( r)$ is $2^{r+4}-2$; thus, the magnetization is 

\begin{equation}
m(3B-4K \le L <  B) = \frac{3}{2}\left(\frac{6}{4^{3}}+\sum_{r=1}^{\infty}\frac{2^{r+4}-2}{4^{r+3}}\right)=\frac{1}{2}.
\end{equation}

\noindent
(b) $2K-B \le L <  3B-4K$

\begin{figure}[h]
\begin{center}
\includegraphics[width=.75\linewidth]{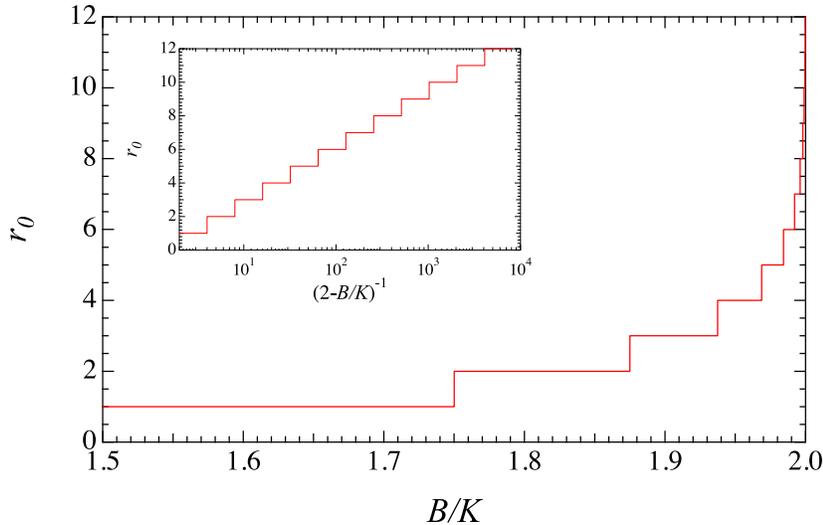}
\end{center}
\caption{(Color online) Number of steps $r_0$ as a function of $B/K\;(=\alpha)$.
The inset shows an enlarged plot of $B/K\simeq 2$. 
}
\label{fig:5}
\end{figure}

First, it should note that the condition $\frac{3K}{2}\le B$, which leads to $2K-B \le 3B-4K$, ensures the existence of the present region of $L$.
We get $a_{3}=32e^{5B}$, $b_{3}=e^{\lambda}a_{3}$, and $c_{3}=e^{2\lambda}a_{3}$, where $\lambda=-2B+4K+2L-\log 2$.
We also obtain the largest term of each $G(r)$, which is given by
\begin{equation}
G(r)=\begin{cases}
2^{r+1} & L < L_{r} \\
2^{r+2}-2 & L \ge L_{r}  ,
\end{cases}
\end{equation}
where
\begin{equation}
   L_{r}=\frac{2^{r}}{2^{r}-1}\left(B-2K+\frac{B}{2^{r}}\right).   
\end{equation}
Here, $L_{r}$ must satisfy the inequality $ 2K-B \le L_{r } <  3B-4K$.  If $L_{r} \ge 2K-B$, then $B \ge \left(2-\frac{1}{2^{r}}\right)K$. 
We put the largest integer of $r$ satisfying $B \ge (2-\frac{1}{2^{r}})K$ with $r_{0}$ under given $B$ and $K$;
then, $L_{r}<2K-B$ for $r \ge r_{0}+1$ and $L_{r} \ge 2K-B$ for $r \le r_{0}$.  
Thus, we have following multiple plateaus:
\begin{equation}
   m(L_{r+1}\le L  < L_{r})=\frac{3}{2}\left(\sum_{s=1}^{r}\frac{2^{s+1}}{4^{s+2}}+\sum_{s=r+1}^{\infty}\frac{2^{s+2}-2}{4^{s+2}}\right),   \  \text {for}   \  r =1,2, \cdots, r_{0}-1,
\end{equation}  
and
\begin{equation}
   m(2K-B \le L <  L_{r_{0}})=\frac{3}{2}\left(\sum_{s=1}^{r_{0}}\frac{2^{s+1}}{4^{s+2}}+\sum_{s=r_{0}+1}^{\infty}\frac{2^{s+2}-2}{4^{s+2}}\right)  .
\end{equation}
We have $r_{0}$ steps of the magnetization in this field region.
When $B=1.8$ and $K$=1, two steps, $m=\frac{17}{64}$ and $\frac{59}{256}$, are obtained. [One step is obtained in region (a) and two steps are obtained in region (b); thus,  we have three steps in total in region (iii).]
When $B \rightarrow2K-0$, then $r_{0} \rightarrow \infty $, and  the number of  steps becomes $\infty$.
In Fig. \ref{fig:5}, we show $r_0$ as a function of $B/K$.

\subsubsection{$0 \le L <2K-B$}\label{sec:3-2-4}

\begin{figure}[h]
\begin{center}
\includegraphics[width=.75\linewidth]{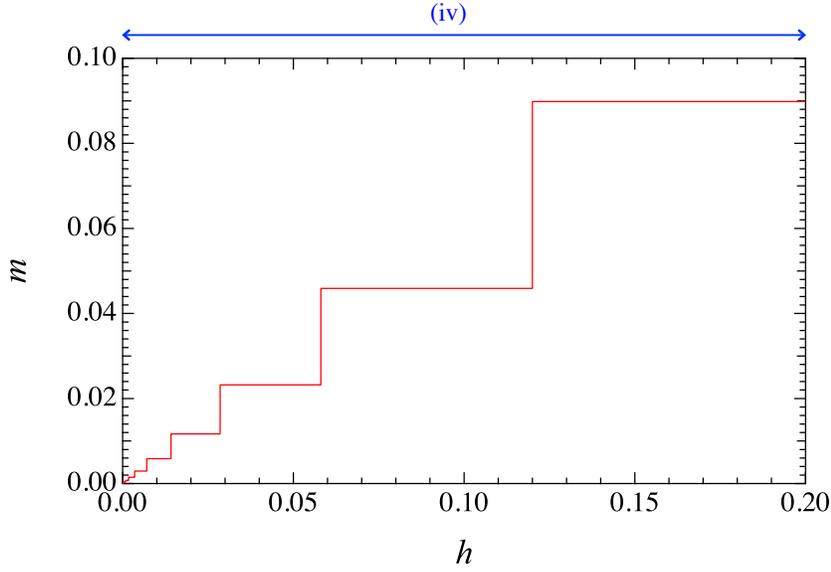}
\end{center}
\caption{(Color online) Magnetization process in region (iv) for $B/K=1.8$. 
 }
\label{fig:6}
\end{figure}

A similar procedure gives Table~{\ref{table:5}}.
From Table~{\ref{table:5}}, the initial value is given by $a_{3}=8e^{B+8K-4L} $ and $\lambda=4L$. 
Thus, $G(r)$ is obtained as 
\begin{equation}
G(r)=\begin{cases}
2^{r+2} & L < L_{r} \\
2^{r+3}-2 & L \ge L_{r}  ,
\end{cases}
\end{equation}
where
\begin{equation}
   L_{r}=\frac{B}{2^{r+1}-1}.
\end{equation}
If we denote the maximum $r$ satisfying $B>\left(2-\frac{1}{2^{r}}\right)K$ for given $B$, and $K$ as $r_{0}$, then $L_{1}>L_{2}> \cdots >L_{r_{0}} >2K-B$. 
Thus, we get
\begin{equation}
   G(r)=\begin{cases}
      2^{r+2} & \mbox{for $r \le r_{0}$} \\
      2^{r+3}-2 & \mbox{for $r>r_{0}$}
   \end{cases}
\end {equation} 
and the magnetization is obtained as
\begin{equation}
   m(L_{r_{0}+1} \le L < 2K-B) = \frac{3}{2}\left(-\frac{4}{4^{2}}+\sum_{s=1}^{r_{0}}\frac{2^{s+2}}{4^{s+2}}+\sum_{s=r_{0}+1}^{\infty}\frac{2^{s+3}-2}{4^{s+2}}\right)
=\frac{3}{2^{r_{0}+3}}-\frac{1}{4^{r_{0}+2}}  ,
\end{equation}
and for $s \ge r_{0}+1$,
\begin{equation}
   m(L_{s+1}\le L < L_{s})=\frac{3}{2^{s+3}}-\frac{1}{4^{s+2}}.
\end{equation}
We also get  the IMS structure. 
When $B/K=1.8$, we have $r_{0}=2, m(L_{3}\le L <B-2K)=\frac{23}{256}\simeq 0.0898438$, $m(L_{4}\le L <L_{3})=\frac{47}{1024}\simeq 0.0458984$, 
$\cdots$ and $m(L <L_{s})  \stackrel{s\rightarrow \infty}{\to} 0$ (see Fig. \ref{fig:6}). 
Applying a small magnetic field on the spin liquid phase leads to an increase in the magnetization, which is roughly linear in $h$.
This is so unique because the discreteness of energy levels of Ising systems tends to prevent a small perturbation from changing any physical quantities.

The spin liquid phase has residual entropy, as shown in Fig. 9 of Ref. 19, and the Zeeman energy takes various values in this degenerate manifold in contrast to the paramagnetic phase.
It is instructive to note that the magnetization $M=\sum_i \sigma_i$ is a conserved quantity in the present Hamiltonian,
and we have, respectively, $M=0$ and $M=N/2$, where $N$ is the total number of vertices, in the paramagnetic ($B/K>2$) and antiferromagnetic ($B/K<1$) phases.
For $1<B/K<2$, the degenerate ground state manifold consists of spin configurations with various $M$ values, which is a necessary condition for the appearance of  the IMS structure.

\begin{table}[t]
\centering
\caption{Most dominant terms in $a_3$, $b_3$, and $c_3$
as a function of $L\in[0,2K-B)$ for $B\in[\frac{3}{2}K,2K)$.}
\label{table:5}
\begin{tabular}{c|c} \hline
$L$ &  $[0,2K-B)$   \\ \hline
$a_{3}$ & $8e^{B+8K-4L}$  \\ \hline
$b_{3}$ & $8e^{B+8K}$  \\ \hline
$c_{3}$ & $8e^{B+8K+4L}$ \\  \hline
\end{tabular}
\end{table}

\subsection{Case of $B=0$}
 
We consider the case of $B=0$, i.e., without frustrating the AF diagonal bond.
The largest terms in $a_2,b_2$, and $c_2$ are given in Table~{\ref{table:6}}.
 
\begin{table}[h]
\centering
\caption{Most dominant terms in $a_2$, $b_2$, and $c_2$
as a function of $L$ for $B=0$.}
\label{table:6}
\begin{tabular}{c|c|c}
\hline
$L$ & $[0,2K)$ & $[2K,\infty)$ \\ \hline
$a_{2}$ & $e^{-2L+4K}$ & $e^{2L-4K}$ \\ \hline
$b_{2}$ &  \multicolumn{1}{c}{$e^{2L}$} &  $e^{2L}$ \\ \hline
$c_{2}$ &  \multicolumn{1}{c}{$e^{2L+4K}$} & $e^{2L+4K}$ \\  \hline
\end{tabular}
\end{table}

\begin{figure}[h]
\begin{center}
\includegraphics[width=.75\linewidth]{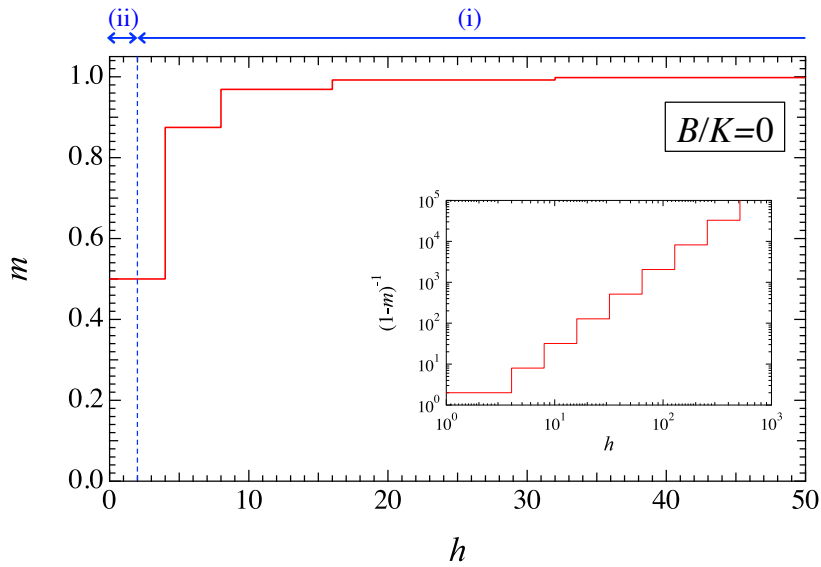}
\end{center}
\caption{(Color online) Magnetization $m$ as a function of $h=L/K$ for $\alpha=B/K=0$.
The inset shows an enlarged plot for the high-field region. 
}
\label{fig:7}
\end{figure}

Here, we show our calculated magnetization curve in Fig.~\ref{fig:7}.
The regions indicated in this figure, (i) $2K\le L<\infty$ and (ii) $0\le L<2K$, correspond to those in Table \ref{table:6}.
We have only the high-field IMS structure around $(m,L)=(1,\infty)$,
which indicates that the other  IMS structure seen in the previous two cases originates from the frustrating AF diagonal bond.
When $B=0$, the system is not frustrated and all the bond energies can be minimized in an antiferromagnetic state where spins on $(2s+d)$-type sites are up and the others are down.
The total number of $(2s+d)$-type sites is 3/4 of the total number of sites, and thus the magnetization per site in the antiferromagnetic ground state at low magnetic fields is 1/2.

We present the calculation process for regions (i) and (ii) in {\it\ref{sec:3-3-1}}-{\it\ref{sec:3-3-2}}, respectively.

\subsubsection{$2K \le L$}\label{sec:3-3-1}

In this case, from Table~{\ref{table:6}}, we get $a_2=e^{2L-4K} $ and $\lambda=4K$. Thus, we get
 
\begin{equation}
   G(r)=\begin{cases}
      -2 &   L  < 2^{r+1}K \\
      2   &  L \ge  2^{r+1}K .
\end{cases}
\end {equation} 
for $r\geq 1$.
Substituting this into Eq.~(\ref{eq:9}), we obtain
\begin{equation}
   m(2^r K \le L < 2^{r+1}K)=1-\dfrac{2}{4^r}.
\end{equation}
The magnetization $m_i$ for $(2^is+d)$-type vertices is shown in Table~\ref{table:6p}.
The absence of AF diagonal bonds makes spins on $(2^is+d)$-type vertices flip in one step as $m_i=-1\to+1$.

Note that we obtain the IMS structure in the high-field region,
although there exists no frustration interaction in the present Hamiltonian.
This high-field IMS structure originates from the competition between the nonfrustrated antiferromagnetic interaction $K$ 
and the magnetic field $L$ and the self-similarity of hierarchical lattices.
As will be shown later, the ferromagnetic interaction case ($K\rightarrow -K$) gives no high-field IMS structure.

\begin{table}[h]
\centering
\caption{Local magnetizations ${m_i}$ as a function of $L$ for $B=0$.}
\label{table:6p}
\begin{tabular}{c|c|c|c|c} \hline
$L$                       & $m_1$ & $m_2$ & $m_3$ & $m_4$   \\ \hline
$[2K,4K)$    &    1       &      -1     &     -1     &     -1       \\
$[4K,8K)$     &    1       &      1     &     -1     &      -1      \\
$[8K,16K)$     &    1       &      1     &      1     &      -1      \\\hline
\end{tabular}
\end{table}

\subsubsection{$0<L<2K$}\label{sec:3-3-2}

In this region, we cannot obtain the simple recursion equations independent of the applied field region.
Thus, we consider a different approach. From Eq. (\ref{eq:6}), we have
\begin{eqnarray}
   e^{-2L}c_n-e^{2L}a_n&&\hspace{-6mm}=(e^{-L}c_{n-1}-e^{L}a_{n-1})(e^{-L}c_{n-1}+e^{L}a_{n-1})
   \nonumber\\
   &&\hspace{20mm}\times(e^{-2L}c_{n-1}^2+2b_{n-1}^2+e^{2L}a_{n-1}^2).
\end{eqnarray}
If $e^{-L}c_{n-1} >e^La_{n-1}$, then we have $e^{-2L}c_n >e^{2L}a_n$, and $e^{-L}c_n > e^{L}a_n$ because $e^{-L}c_n >e^{3L}a_n >e^La_n$. 
Since $e^{-L}c_2 > e^La_2$, we obtain the above results by the inductive method. 
Inserting these results into Eq.~(\ref{eq:6}), we obtain $b_n=e^{-2L}c_{n-1}^2b_{n-1}^2$.
From Eq.~(\ref{eq:6}) $c_n=(e^Lb_{n-1}^2+e^{-L}c_{n-1}^2)^2 \ge 4b_{n-1}^2c_{n-1}^2 >e^{2L}b_n > e^L b_n$.
We insert this into Eq.~(\ref{eq:6}) and get
\begin{equation}
   c_n=e^{-2L}c_{n-1}^4,\;\;\mbox{and}\;\;\;Z_n=e^{-2L}c_n.
\end{equation}
Thus, we get 
\begin{equation}
   m(0<L<2K)=\dfrac{3}{2}\left(\dfrac{2}{4}-2\sum_{r=2}^{\infty}\dfrac{1}{4^r}\right)=\dfrac{1}{2}.
\end{equation}
Our diamond hierarchical lattice is bipartite at $B=0$.
The up sublattice is constituted by $(2s+d)$-type vertices and the down sublattice is constituted by the other vertices.
Because the total number of sites in the up sublattice is three times as large as that in the down sublattice, our antiferromagnetic state for $0<L<2K$ is ferrimagnetic and has  spontaneous magnetization.

\section{Magnetization of Ferromagnet with Frustration}\label{sec:4}
We consider the ferromagnetic Hamiltonian with frustration, which is given by
\begin{equation} 
   {\cal H}=-J\sum_{\langle i,j\rangle} \sigma_i \sigma_j +\alpha J \sum_{\langle\!\langle i,j\rangle\!\rangle}\sigma_i \sigma_j -H\sum_{i}\sigma_{i}.
\end{equation}
The ferromagnetic exchange interaction does not compete with the Zeeman interaction, in contrast to the previous antiferromagnetic case,
and thus we obtain the saturated magnetization at a finite saturation field, as shown below.

\subsection{Case of $2K \le B <3K$}
The zero-field phase for this case is the paramagnetic phase.\cite{Kobayashi2009}
As the initial values for the recursion formulas, we have Table~{\ref{table:7}}.

\begin{table}[h]
\centering
\caption{Most dominant terms in $a_2$, $b_2$, and $c_2$
as a function of $L$ for $B\in[2K,3K)$.}
\label{table:7}
\begin{tabular}{c|c|c|c|c}
\hline
$L$ & $[0,B-2K)$ &$[B-2K,B)$ & $[B,B+2K)$  & $[B+2K,\infty)$  \\ \hline
$a_{2}$ & $2e^{B}$ & \multicolumn{1}{c}{$e^{2L-B+4K}$} &  \multicolumn{1}{c}{$e^{2L-B+4K}$}& $e^{2L-B+4K}$ \\ \hline
$b_{2}$ &  \multicolumn{1}{c}{$2e^{B}$} & $2e^{B}$&   \multicolumn{1}{c}{$e^{2L-B}$}& $e^{2L-B}$  \\  \hline
$c_{2}$ &   \multicolumn{1}{c} {$2e^{B}$}&  \multicolumn{1}{c}{$2e^{B}$} &$2e^{B}$  & $e^{2L-B-4K}$ \\ \hline
\end{tabular}
\end{table}

\begin{figure}[h]
\begin{center}
\includegraphics[width=.75\linewidth]{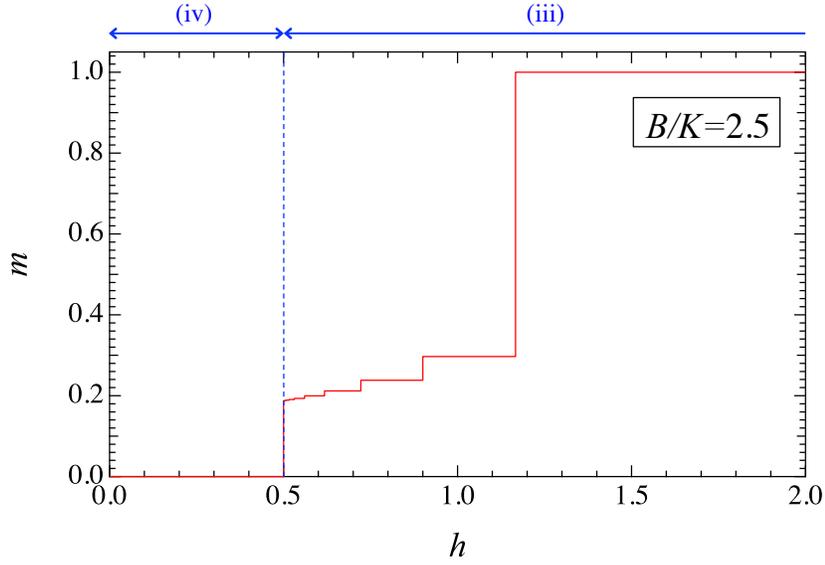}
\end{center}
\caption{(Color online) Magnetization $m$ as a function of $h=L/K$ for $\alpha=B/K=2.5$.
Regions (i) and (ii), which are in $h>B/K=2.5$, are not shown in this figure. 
}
\label{fig:8}
\end{figure}

Typical features of the resultant magnetization curves are as follows.
We show a calculated result in Fig.~\ref{fig:8}, where $B/K=2.5$ is chosen.
In Table~\ref{table:7}, there exist four regions, (i) $2K+B\le L<\infty$, (ii) $B\le L<2K+B$, (iii) $B-2K\le L<B$, and (iv) $0\le L<B-2K$.
The last two are indicated in Fig.~\ref{fig:8}.
In region (iii), we have the finite saturation field $L=\frac{3B-4K}{3}$, which differs from the antiferromagnetic case. 
Also, we find that the IMS structure appears around $(m,L)=(\frac{3}{16},L_{\rm c})$ with $L_{\rm c}=B-2K$.
In region (iv), the magnetization vanishes, because strong AF diagonal bonds form $(\uparrow,\downarrow)$ pairs.

We present the calculation process for regions (i)-(iv) in {\it\ref{sec:4-1-1}}-{\it\ref{sec:4-1-4}}, respectively.

\subsubsection{$B+2K \le L$}\label{sec:4-1-1}

From Table~{\ref{table:7}}, we get $a_{2}=e^{4K-B+2L}$ and $\lambda=-4K$. Thus, $G(r)=2$ for all $r$.
We immediately obtain
\begin{equation}
   m(B+2K\le L)=\frac{3}{2}\left(\frac{1}{2}+\sum_{r=1}^{\infty} \frac{2}{4^{r+1}}\right)=1.
\end{equation}

\subsubsection{$B\le L < B+2K$}

Using the recursion formula, we get $a_3$, $b_3$, and $c_3$ in Table~{\ref{table:8}}.
From Table~{\ref{table:8}}, we get $a_{3}=e^{-5B+16K+10L}$, $\lambda=-8K$, and $G(r)=2$. The magnetization is $m=1$.

\begin{table}[h]
\centering
\caption{Most dominant terms in $a_3$, $b_3$, and $c_3$
as a function of $L\in[B,B+2K)$ for $B\in[2K,3K)$.}
\label{table:8}
\begin{tabular}{c|c}
\hline
$L$ &  $[B,B+2K)$   \\ \hline
$a_{3}$ &  $e^{-5B+16K+10L}$  \\ \hline
$b_{3}$ &  $e^{-5B+8K+10L}$  \\ \hline
$c_{3}$ &  $e^{-5B+10L}$  \\ \hline
\end{tabular}
\end{table}

\subsubsection{$B-2K \le L < B $}

We obtain the following table in this region using the recursion formula twice.

\begin{table}[h]
\centering
\caption{Most dominant terms in $a_3$, $b_3$, $c_3$, $a_4$, $b_4$, and $c_4$
as a function of $L\in[B-2K,B)$ for $B\in[2K,3K)$.}
\label{table:9}
\begin{tabular}{c|c|c|c}
\hline
$L$ &   $[B-2K,\frac{3B-4K}{3})$   &    $[\frac{3B-4K}{3},B-K)$                  & $[B-K,B)$                                 \\ \hline
$a_{3}$ &  $8e^{B+8K+4L}$         &    \multicolumn{1}{c}{$e^{-5B+16K+10L}$}                       &  $e^{-5B+16K+10L}$             \\ \hline
$b_{3}$ &  \multicolumn{1}{c}{$16e^{3B+4K+2L}$}     &    $16e^{3B+4K+2L}$                         &  $4e^{-B+8K+6L}$                     \\ \hline
$c_{3}$ &  \multicolumn{1}{c}{$32e^{5B}$}                  &    \multicolumn{1}{c}{$32e^{5B}$}                                     &  $32e^{5B}$                             \\ \hline
$a_{4}$ &                                      &     \multicolumn{1}{c}{$e^{-21B+64K+42L}$}                    &   $e^{-21B+64K+42L}$             \\ \hline
$b_{4}$ &                                      &     $16^2e^{-5B+40K+26L}$               & $16e^{-13B+48K+34L}$         \\ \hline
$c_{4}$ &                                      &     $16^4e^{11B+16K+10L}$               & $16^2e^{-5B+32K+26L}$       \\ \hline
\end{tabular}
\end{table}

\noindent
(a) $B-K\le L < B$

From Table~{\ref{table:9}}, we have $a_{4}=e^{-21B+64K+42L} $ and $\lambda=8B-16K-8L+4\log 2$, and $G(r)=2$ for all $r$.
Thus, the magnetization is
\begin{equation}
   m(B-K \le L < B)=1. 
\end{equation}

\noindent
(b) $\frac{3B-4K}{3} \le L < B-K$

From Table~{\ref{table:9}}, we get $a_4=e^{-21B+64K+42L}$ and $\lambda=16B-24K-16L+8\log 2$. 
Thus, $G(r)=2$ for all $r$. The magnetization is 
\begin{equation}
   m\left(\dfrac{3B-4K}{3} \le L < B-K\right)=1. 
\end{equation}
The saturation field is $L=\frac{3B-4K}{3}$.

\noindent
(c) $B-2K\le L < \frac{3B-4K}{3}$

From Table~{\ref{table:9}}, we get $a_{3}=8e^{B+8K+4L}$ and $\lambda=2B-4K-2L+\log 2$.
If $L_{r}$ is defined by
\begin{equation} 
   L_{r}=B-\frac{2^{r+1}}{2^{r}+1}K,
\end{equation}
then we obtain 
\begin{equation}
G(r)=\begin{cases}
-2^{r+1} & L < L_{r} \\
 2 & L \ge L_{r}  
\end{cases}.
\end{equation}
The magnetization is given by 
\begin{equation}
   m(L_{r+1} \le L < L_{r})=\frac{3}{2}\left(\frac{4}{4^2}-\sum_{s=1}^r\frac{2^{s+1}}{4^{s+2}}+2\sum_{s=r+1}^{\infty}\frac{1}{4^{s+2}}\right)
=\frac{3}{16}+\frac{3}{2^{r+4}}+\frac{1}{4^{r+2}}.
\end{equation}
Since $L_{\infty}=B-2K$ and $m(L=L_{\infty})=\frac{3}{16}$, we have the  IMS structure. 
The step width of the magnetization is independent of $B$ and $K$, showing universality, but the field intensity at the jump depends on $B$ and $K$. 
Some magnetizations, when $B/K=2.5$, $m(0.9\le L<1.333)=\frac{19}{64}=0.296875$, 
$m(0.7222\le L<0.9)=\frac{61}{256}\simeq0.238281$, $m(0.6176\le L<0.7222)=\frac{217}{1024}\simeq0.211914$, are shown in Fig.~\ref{fig:8}.

\subsubsection{$L<B-2K$}\label{sec:4-1-4}

In this case, we have $a_{2}=b_{2}=c_{2}=2e^{B}$ and $\lambda=0$. 
Thus, we immediately obtain $m=0$.
The vanishing of the magnetization is due to the formation of $(\uparrow,\downarrow)$ pairs by the AF diagonal bonds.

\subsection{Case of $\frac{5K}{3}  \le  B < 2K$}

\begin{figure}[h]
\begin{center}
\includegraphics[width=.75\linewidth]{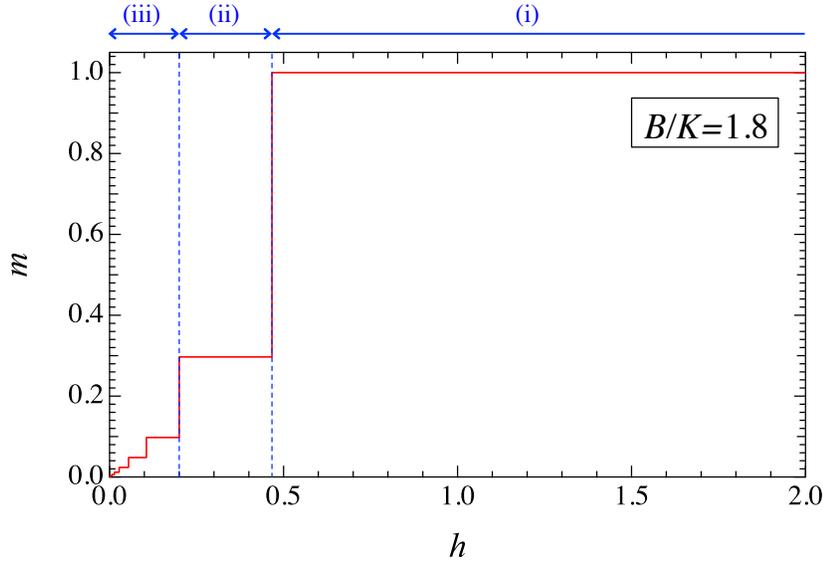}
\end{center}
\caption{(Color online) Magnetization $m$ as a function of $h=L/K$ for $\alpha=B/K=1.8$. 
}
\label{fig:9}
\end{figure}

The zero-field ground state for $\frac{5K}{3}\le B<2K$ is the spin liquid state.\cite{Kobayashi2009}
Here, we describe a typical feature of the magnetization curve.
In Fig. \ref{fig:9}, we show the calculated result with $B/K=1.8$.
The regions indicated in Fig.~\ref{fig:9} are (i) $\frac{3B-4K}{3}\le L<\infty$, (ii) $2K-B\le L<\frac{3B-4K}{3}$, and (iii) $0\le L<2K-B$.
In region (i), we have a saturation field $L=\frac{3B-4K}{3}$.
In region (ii), the total number of magnetization plateaus depends on $B/K$. 
In region (iii), the IMS structure appears.
We can observe that a small magnetic field induces a small magnetization.

We present the calculation process for regions (i)-(iii) in sects.{\it\ref{sec:4-2-1}}-{\it\ref{sec:4-2-3}}, respectively.

\subsubsection{$\dfrac{3B-4K}{3}  \le L $}\label{sec:4-2-1}

In this region, a similar calculation leads to the result that the  magnetization is 
\begin{equation}
   m\left(\dfrac{3B-4K}{3} \le L\right) = 1.
\end{equation}
The saturation field is given by $L=\frac{3B-4K}{3}$.

\subsubsection{$2K-B \le L <\dfrac{3B-4K}{3}$}

For $r\le r_0$, where $r_0$ is the largest integer that satisfies $L_r=B-\frac{2^{r+1}}{1+2^r}K >2K-B$, $G(r)$ is given by  
\begin{equation}
G(r)=\begin{cases}
-2^{r+1} & L < L_{r} \\
 2 & L \ge L_{r}  
\end{cases}.
\end{equation}
The magnetization is 
\begin{equation}
   m\left(2K-B \le L <L_{r_0}\right)=\dfrac{3}{16}+\dfrac{3}{2^{r_0+4}}+\dfrac{1}{4^{r_0+2}},
\end{equation}
\begin{equation}
   m\left(L_{r+1}\le L <L_r\right)=\dfrac{3}{16}+\dfrac{3}{2^{r+4}}+\dfrac{1}{4^{r+2}}\;\;\;\;\;\mbox{for $r<r_0$}.
\end{equation}
We have $r_{0}$ steps of the magnetization in this field region.
When $B=1.8$ and $K$=1, one step, $m=\frac{19}{64}$, is obtained (see Fig.~{\ref{fig:9}}).
When $B \rightarrow2K-0$, then $r_{0} \rightarrow \infty $, and  the number of  plateaus  becomes $\infty$.

\subsubsection{$0 \le L <2K-B$}\label{sec:4-2-3}

We have $a_{3}=8e^{B+8K+4L}$, and $\lambda=-4L$, and so we also get
\begin{equation}
G(r)=\begin{cases}
-2^{r+2} & L < L_{r} \\
 2 & L \ge L_{r}  
\end{cases}.
\label{eq:Gr}
\end{equation}
for $r\ge r_{0}$, where $r_{0}$ is the smallest integer in $r$ satisfying $L_{r}=\dfrac{B}{2^{r+1}+1} < 2K-B$, and
Eq. (\ref{eq:Gr}) gives the magnetization:
\begin{equation}
m(L_{r_{0}}\le L <2K-B) =\dfrac{3}{2^{r_{0}+2}}+\dfrac{1}{4^{r_{0}+1}},
\end{equation}
\begin{equation}
m(L_{r+1} \le L < L_{r} ) = \dfrac{3}{2^{r+3}}+\dfrac{1}{4^{r+2}}\;\;\;\;\mbox{for $r \ge r_{0}$} .
\end{equation}
Thus, we also obtain the  IMS structure (see Fig. \ref{fig:9}).

\subsection{Case of $B=0$}

When $L \ge 2K$, the largest terms are respectively $a_2=e^{4K+2L}$, $b_2=e^{-4K}a_2$, and $c_2=e^{-8K}a_2$. 
Thus, an initial term in Eq.~(\ref{eq:9}) is $a_2$, and $\lambda=-4K$. 
We get $m(2K \le  L)=1$.

For $0<L<2K$, it can be proved that $b_n=e^{-2^n K}a_n$ and $c_n\le e^{-2^n L}a_n$.
A proof by induction is as follows.
From Eq.~(\ref{eq:3}), we obtain $b_2=e^{-4K}a_2$ and $c_2=e^{-4L}a_2$, which shows that the two relations hold when $n=2$.
For $n>2$, assuming $b_n=e^{-2^n K}a_n$ and $c_n\le e^{-2^n L}a_n$, we find that Eq.~(\ref{eq:6}) gives
\begin{equation}
   a_{n+1}=e^{2L}a_n^4,
\end{equation}
\begin{equation}
   b_{n+1}=e^{2L}a_n^2b_n^2=e^{-2^{n+1}K}a_{n+1}
\end{equation}
and
\begin{equation}
   c_{n+1}\le(e^{-2^{n+1}K}+e^{-2L-2^{n+1}L})^2a_{n}^4e^{2L}<(e^{-2^{n}L}+e^{-2L-2^{n+1}L})^2a_{n+1}=e^{-2^{n+1}L}a_{n+1},
\end{equation}
where we have used the relation $e^{-2^{n+1}K}<e^{-2^{n}L}$. Therefore, the two relations hold for any $n$.
Because of $b_n=e^{-2^n K}a_n$ and $c_n\le e^{-2^n L}a_n$, we immediately have $a_n>b_n$ and $a_n>c_n$.
Then, the partition function $Z_n$ in Eq.~(\ref{eq:5}) can be written as $Z_n=e^{2L}a_n$. 
We put $a_{n}=e^{2L}a_{n-1}^4$ into this equation for $Z_n$, then we find $m(0<L<2K)=1$. 

Kobayashi {\it et al.} showed for $0 \le B \le  1$ and $L=0$ that the ground state is ferromagnetic.\cite{Kobayashi2009}
Thus when $B=0$, $m(0 \le  L)=1$ for all external fields.

\section{Conclusions}\label{sec:5}

The magnetizations for both  ferro- and antiferromagnets with the frustration on the diamond hierarchical lattice at absolute zero are exactly obtained. 
Our lattice, which contains vertices with high coordination numbers, has an intrinsic long-range nature, to which
the appearance of the IMS structure can be ascribed.

In the antiferromagnet of the unlimited system, the saturated magnetization cannot be realized under finite magnetic fields.
At high magnetic fields, the spin flip begins in a small type of vertex and moves to the large ones; spins at the vertices of the $(2^{s}s+d)$ type flip upwards, whenever magnetic fields are added to $(2^{s}K+B)$.
Even if $B=0$, namely, no frustration interactions are included in the Hamiltonian, we get this IMS structure around $(m,h)=(1,\infty)$.
On the other hand, we have a saturated magnetization in the ferromagnet. 
These facts show that the competition between a nonfrustrated antiferromagnetic interaction and the magnetic field is highlighted by the long-range nature of hierarchical lattices.

A frustrated AF diagonal bond makes the high-field magnetization process somewhat complicated; a two-step spin flip of each vertex occurs.
Also, the AF diagonal bond gives an additional IMS structure in the low-field region in both ferro- and antiferromagnetic cases.
Applying a small magnetic field on the spin liquid phase gives a small magnetization; the IMS structure around $(m,h)=(0,0)$ appears.
For the paramagnetic phase, applying a small magnetic field does not give any magnetization, and thus there exists a threshold $h_{\rm c}$ for getting a nonzero magnetization.
In this case, we have an infinitely short plateau starting from the threshold: the IMS structure around $(m,h)=(m(h_{\rm c}+0)=\frac{3}{16},h_{\rm c})$.

As for the change in the lowest-energy spin configurations as a function of magnetic field, we have succeeded in elucidating it only in a high-magnetic-field region.
The issue of low-field spin configurations remains a future problem.


\end{document}